\documentclass[12pt]{article}
\usepackage{epsfig}
\begin{document}

\baselineskip=14.2pt
\pagestyle{plain}
\setcounter{page}{1}

\renewcommand{\thefootnote}{\fnsymbol{footnote}}

\newcommand{\nc}{\newcommand}
\nc{\grad}{\nabla}
\nc{\tr}{\mathop{\rm tr}}
\nc{\half}{{1\over 2}}
\nc{\third}{{1\over 3}}
\nc{\be}{\begin{equation}}
\nc{\ee}{\end{equation}}
\nc{\bea}{\begin{eqnarray}}
\nc{\eea}{\end{eqnarray}}

\def\R{{\bf R}}
\def\Tr{{\rm Tr}}
\nc{\dint}[2]{\int\limits_{#1}^{#2}}
\nc{\D}{\displaystyle}
\nc{\PDT}[1]{\frac{\partial #1}{\partial t}}
\nc{\tw}{\tilde{w}}
\nc{\tg}{\tilde{g}}
\nc{\newcaption}[1]{\centerline{\parbox{5.6in}{\caption{#1}}}}
\def\href#1#2{#2} 

\nc{\al}{\alpha}
\nc{\ga}{\gamma}
\nc{\de}{\delta}
\nc{\ep}{\epsilon}
\nc{\ze}{\zeta}
\nc{\et}{\eta}
\renewcommand{\th}{\theta}
\nc{\Th}{\Theta}
\nc{\ka}{\kappa}
\nc{\la}{\lambda}
\nc{\rh}{\rho}
\nc{\si}{\sigma}
\nc{\ta}{\tau}
\nc{\up}{\upsilon}
\nc{\ph}{\phi}
\nc{\ch}{\chi}
\nc{\ps}{\psi}
\nc{\om}{\omega}
\nc{\Ga}{\Gamma}
\nc{\De}{\Delta}
\nc{\La}{\Lambda}
\nc{\Si}{\Sigma}
\nc{\Up}{\Upsilon}
\nc{\Ph}{\Phi}
\nc{\Ps}{\Psi}
\nc{\Om}{\Omega}
\nc{\ptl}{\partial}
\nc{\del}{\nabla}
\nc{\ov}{\overline}
\nc{\gsl}{\!\not}
\nc{\bi}[1]{\bibitem{#1}}
\nc{\fr}[2]{\frac{#1}{#2}}
\nc{\gm}{\mbox{$\gamma_{\mu}$}}
\nc{\gn}{\mbox{$\gamma_{\nu}$}}
\nc{\Le}{\mbox{$\fr{1+\gamma_5}{2}$}}
\nc{\Ri}{\mbox{$\fr{1-\gamma_5}{2}$}}
\nc{\GD}{\mbox{$\tilde{G}$}}
\nc{\gf}{\mbox{$\gamma_{5}$}}
\nc{\Ima}{\mbox{Im}}
\nc{\Rea}{\mbox{Re}}
\def\lasq{\langle 0| {\rm Tr}\, \lambda^2 |0\rangle}
\def\lasqd{\langle 0| {\rm Tr}\, \lambda^2 (x)\, {\rm Tr}\, \lambda^2(x')|0\rangle}
\def\laph{\langle 0| {\rm Tr}\, \lambda^2 (x)\, \varphi^2 (x') |0\rangle}
\def\phph{\langle 0| \varphi^2 (x)\, \varphi^2 (x') |0\rangle}
\def\phsq{\langle 0| {\rm Tr}\, \phi^2|0\rangle}
\def\phsqb{\langle 0| \varphi^2|0\rangle}
\nc{\av}{\langle \ph\rangle}


\def\Z{{\bf Z}}
\def \ci {\cite}
\def \foot {\footnote}
\def \bi{\bibitem}
\nc{\rf}[1]{(\ref{#1})}
\def \del{\partial}
\def \lab {\label}
\def \ha {\textstyle{1\ov 2}}

\def \np { Nucl. Phys. }
\def \pl { Phys. Lett. }
\def \mpl { Mod. Phys. Lett. }
\def \prl { Phys. Rev. Lett. }
\def \pr  { Phys. Rev. }
\def \cqg { Class. Quantum Grav.}
\def \jmp { J. Math. Phys. } 
\def \ap { Ann. Phys. }
\def \ijmp { Int. J. Mod. Phys. }


\begin{titlepage}

\begin{flushright}
TPI-MINN-99/43\\
UMN-TH-1817-99\\
hep-th/9909073 
\end{flushright}
\vfil

\begin{center}
{\LARGE Instantons at Strong Coupling, Averaging over Vacua, 
 and the Gluino Condensate
}
\end{center}

\vfil
\begin{center}
{ Adam Ritz and 
  Arkady Vainshtein}\\
\vspace{2mm}
{\em Theoretical Physics Institute, University of Minnesota,
 MN 55455, USA \\
 {\tt aritz@mnhepw.hep.umn.edu, vainshtein@mnhep1.hep.umn.edu}}
\vspace{3mm}
\end{center}

\vfil

\begin{center}
{\large Abstract}
\end{center}

\noindent
We consider instanton contributions to chiral correlators,
such as $\langle 0| {\rm Tr} \lambda^2 (x) {\rm Tr} \lambda^2(x')|0\rangle$,
in ${\cal N}\!\!=\!1$ supersymmetric Yang-Mills theory with either 
light adjoint or
fundamental matter.  Within the former model, extraction of the gluino
condensate from a connected {\mbox 1-instanton} diagram, evaluated
at strong coupling, can be contrasted with expectations from
the Seiberg-Witten solution perturbed to an ${\cal N}\!\!=\!1$ vacuum. 
We observe a numerical discrepancy, coinciding with that observed 
previously in ${\cal N}\!\!=\!1$ SQCD. Moreover, since knowledge of the vacuum 
structure is complete for softly broken ${\cal N}\!\!=\!2$ Yang-Mills, 
this model 
serves as a counterexample to the hypothesis of Amati {\em et al.} that 
1-instanton calculations at strong coupling can be interpreted 
as averaging over  vacua. Within ${\cal N}\!\!=\!1$ SQCD, we point out
that the connected contribution to the relevant correlators
actually vanishes in the weakly coupled Higgs phase, 
despite having a nonzero value through infra-red effects when calculated 
in the unbroken phase.

\vfil
\begin{flushleft}
September 1999
\end{flushleft}
\end{titlepage}


\newpage

\renewcommand{\thefootnote}{\arabic{footnote}}
\setcounter{footnote}{0}

\section{Introduction}

Historically, the first arena in which supersymmetry and 
nonperturbative instanton effects were successfully integrated was in 
the direct
calculation of certain chiral correlators~\cite{nsvz1}. These 
correlators involve the lowest components of chiral superfields, and
a well-known example in ${\cal N}\!\!=\!1$ supersymmetric Yang-Mills (SYM)
with gauge group SU(2) is the 2-point function of 
gluino bilinears $\lasqd$. Chirality 
ensures that there can be no contribution at any order in 
perturbation theory, while at first sight it is also highly
nontrivial that instantons can lead to a nonzero result without
breaking supersymmetry. That they do, reflects one of the profound features
of supersymmetric gauge dynamics, in that such correlators are 
essentially topological. In flat space, supersymmetry demands 
that they be spacetime independent constants~\cite{nsvz1}, while more
generally, when evaluated in nontrivial geometries, they are
independent of the background metric, and reflect only
specific topological features of the background \cite{svunp}. 
This was made particularly clear by Witten's
subsequent construction of topological field theories \cite{tqft}.

The qualitative picture for instanton generation of these
chiral correlators, and the associated superpotentials, has been clear
for some time. However, the issue of the precise numerical coefficients
has been rather more controversial.  During the eighties, two different
techniques for instanton calculations in this context were developed.
The ``strong-coupling instanton'' (SCI) approach \cite{nsvz1,sc,fs,scmass}
(see \cite{screv} for a review), applies in strongly coupled
theories such as ${\cal N}\!\!=\!1$ SYM where the instanton is an exact 
solution
to the equations of motion, and involves a direct integration
over the instanton zero modes. For a given correlator, only a given
instanton order can contribute, and the integration over instanton
size $\rh$ is peaked at $\rh \sim |x-x'|$. Since
the result is spacetime independent as a consequence of
supersymmetry, by taking $|x-x'|\rightarrow 0$
one expects the calculation to be well defined, and thus the result appears,
mathematically at least, to be ``exact''. The loophole is that although
small size instantons are under control, there is no guarantee that
additional fluctuations at some larger scale, e.g. $\rh\sim \La$, cannot
contribute.

An additional puzzle concerns cluster decomposition. At large distances
$|x-x'|\rightarrow\infty$ one expects the result to factorise. 
Returning to the
SU(2) example, we expect $\lasqd = \lasq^2$,
implying a nonzero gluino condensate. The double valued nature of 
$\lasq$ is consistent with the Witten index \cite{index} for this theory, but 
the puzzle lies in the fact that instantons cannot contribute to
$\lasq$ directly as ${\rm Tr}\,\la^2$ can only saturate two of four
fermionic zero modes. Thus in the instanton approximation 
$\lasq=0$, and one apparently has a violation of cluster decomposition.
A possible resolution to this puzzle, known as the 
``vacuum averaging hypothesis'' was proposed by Amati {\it et al.}
\cite{screv}, within which the SCI calculation reflects an
average over the vacua of the theory. This nicely 
explains why $\lasqd$ for example is nonzero, 
while a direct instanton calculation of $\lasq$ at strong coupling 
necessarily vanishes, since (for gauge group SU(2)) one
obtains contributions of opposite sign from the two chirally 
asymmetric vacua.

The second approach to instanton calculations, known as the
``weak-coupling instanton'' (WCI) technique \cite{ads,wc} (for a 
recent review, see \cite{svrev}),
first makes use of additional fundamental matter fields to put the system at 
weak coupling where the instanton calculation should indeed be 
reliable. The WCI calculation
produces a gluino condensate with a square-root dependence on the
mass $m$ of the matter fields, 
$\langle {\rm Tr} \lambda^2 \rangle \sim \sqrt{m}$.
The important feature, proven in \cite{hol}, is that the mass dependence
observed at weak coupling is exact. The proof is based on supersymmetric
Ward identities \cite{sc}
which demand that $\langle {\rm Tr} \lambda^2 \rangle$ 
be a holomorphic function of $m$,
i.e. $\ptl\, \langle {\rm Tr} \lambda^2 \rangle/\partial \ov{m} =0$, and
on the relation $\ptl \, \langle {\rm Tr} \lambda^2 \rangle/\partial m
=\langle {\rm Tr} \lambda^2 \rangle/2 m$. Therefore we can then take the
limit $m\rightarrow \infty$ to return to strong coupling and 
compare directly with the SCI calculation.

Comparison of the SCI and WCI results at strong coupling leads to
a well-known numerical discrepancy,
\be
\left\langle 0|\, {\rm Tr}\, \lambda^2\, |0\right\rangle_{\rm SC}^2
 \, = \, \frac{4}{5}\, \left\langle 0|\, {\rm Tr}\, \lambda^2\, 
 |0\right\rangle_{\rm WC}^2\,.\label{disc}
\ee
Note that in contrast to the suggestion of vacuum averaging within
the SCI calculation, the WCI calculation necessarily reflects the 
contribution from a single vacuum, as one analytically continues from 
weak coupling where the vacuum state is unambiguous. However, if we assume
that this model has two chirally asymmetric vacua, as implied by the
Witten index \cite{index}, these contribute equally to 
$\lasq_{SC}^2$ and thus the factor of $4/5$ in (\ref{disc}) represents
an incompatibility, the origin of which has been the source of some debate
over the last ten years.

The WCI calculations have the advantage that they are performed at weak
coupling where the relevant contributions are well understood. 
Furthermore, the constraints of
holomorphy have since been used by Seiberg to unearth a wealth of
nonperturbative information about ${\cal N}\!\!=\!1$ gauge 
theories~\cite{seiberg},
and similar tools were also used by Seiberg and Witten in the
context of ${\cal N}\!\!=\!2$ SYM \cite{sw}. Given these successes of 
the general
weak coupling approach, suspicion naturally falls first on the 
strong coupling calculation and it is of interest to know whether
the discrepancy in (\ref{disc}) has any physical content. In 
this regard, the vacuum averaging
hypothesis has been utilised by Kovner and Shifman \cite{ks}, who 
suggested that the numerical discrepancy in (\ref{disc}) could be 
explained in this way within pure ${\cal N}\!\!=\!1$ SYM by 
assuming the existence of 
an additional chirally symmetric vacuum, where $\lasq=0$. Such a vacuum 
must have rather unusual properties and there has been considerable 
debate as to its feasibility \cite{sym}.

In this paper, we are not  concerned with the chirally symmetric vacuum
specifically, but in the underlying vacuum-averaging hypothesis,
and consequently on the interpretation of the SCI calculation itself. Our
main point will be that there is a clear counter-example to this
hypothesis, namely ${\cal N}\!\!=\!1$ SYM with light adjoint matter (or 
in other words softly broken
${\cal N}\!\!=\!2$ SYM). The additional insight necessary to make 
this conclusion
arises by perturbing the Seiberg-Witten solution \cite{sw}, so that the
exact ${\cal N}\!\!=\!1$ vacuum structure may be deduced. 
As is well-known, one finds
only two chirally asymmetric vacua. This result will be briefly
reviewed in Section~2. Knowledge of these vacua allows a simple deduction
of the gluino condensate via use of the Konishi anomaly \cite{ka}.
For comparison, we then consider the direct SCI calculation of
$\lasqd_{SC}$ in this model, finding exactly the same numerical
discrepancy as in Eq.~(\ref{disc}). This failure of vacuum averaging
leaves the interpretation of the SCI calculation rather unclear. 
We discuss this issue in relation to the known quantum 
corrections to the ${\cal N}\!\!=\!2$ moduli space in Section~3.

We turn in Section~4 to ${\cal N}\!\!=\!1$ SQCD, i.e. involving light
fundamental matter, and review several aspects of the WCI and
SCI calculations. In this case we have greater control over the WCI
calculation, for the correlator $\laph$ (where 
$\varphi$ is a squark modulus) in particular, and we
show how factorisation occurs naturally at weak coupling, where
the SCI result (which is a connected diagram) actually
gives no contribution. We also revisit an additional inconsistency 
(and a suggested resolution) of the SCI calculation which arises at 
1-instanton order, in that
the numerical discrepancy between the SCI and WCI calculations for
$\lasqd$ and $\laph$ actually differs, violating the 
Konishi relation \cite{ka}.

In regard to inconsistencies of the SCI calculation in ${\cal N} \!\!=\!1$ 
SQCD, 
we note that recent work on correlators saturated at strong coupling by 
multi-instanton configurations \cite{hklm} has suggested that 
cluster-decomposition may not hold within the SCI approach. 
We briefly comment on the distinction between these results and 
1-instanton effects at the end of Section~4. 
Finally, we conclude in Section~5 with 
some additional remarks on the interpretation of the strong coupling 
calculation.

\section{Exact Vacuum Structure vs SCI Calculations}

In this section we shall present two different calculations of
the gluino condensate within softly broken ${\cal N}\!\!=\!2$ SYM, one 
from the
Seiberg-Witten solution of the ${\cal N}\!\!=\!2$ theory, and the other
from a direct SCI calculation. We find a discrepancy advertised in 
the previous section, which we shall elaborate on in Section~3.

In terms of ${\cal N}\!\!=\!1$ multiplets, softly broken ${\cal N}\!\!=\!2$ 
SYM possesses one vector,
and one adjoint chiral multiplet, the latter having 
mass $m\ll\La$, where $\La$ is the dynamical scale in ${\cal N}\!\!=\!2$ SYM.
Throughout we shall take the gauge group to be SU(2). The gluino
$\la_{\al}$ is the lowest component of the field strength $W_{\al}$
of the real vector superfield $V$, while we shall denote 
the lowest component of the adjoint chiral
superfield $\Ph$, by $\ph$.

The gluino condensate, on which we will focus, is related to 
condensates of scalar fields by  the Konishi anomaly~\cite{ka}. This
anomaly is expressed in terms of the following  operator relation 
\be
 Z_{\Ph}\,\ov{D}^2(\ov{\Ph}\, e^V \Ph)\,  
    = \, 4\Ph\, \frac{\ptl {\cal W}(\Ph)}{\ptl \Phi} + 
     \frac{T_R}{2\pi^2}\,{\rm Tr}\,W^2 \,, \label{kreln}
\ee
where ${\cal W}(\Ph)$ is the ${\cal N}\!\!=\!1$ superpotential, 
$R$ denotes the representation of $\Ph$, and the group factors
for SU(2) are $T_{\rm fund}=1/2$, $T_{\rm ad}=2$. $Z_{\Ph}$ is the
bare field normalisation factor, and note that summation over the colour
components of $\Phi$ is implied. 
An average of the left-hand side of
this equation
over a supersymmetric vacuum  vanishes leading to the following
relation  for the gluino condensate,
\be
\left\langle 0|\, {\rm Tr}\, \lambda^2\, |0\right\rangle \, = \, 
\frac{8\pi^2}{T_R} \,
      \left\langle 0\left|\, \ph\, 
    \frac{\ptl {\cal W}(\ph)}{\ptl \phi}\, \right|0\right\rangle\,.
\label{kanom1}
\ee
In particular, when $\Phi$ is in the adjoint representation of SU(2),
it is natural to choose an ${\cal N}\!\!=\!2$ normalisation for $\Ph$
so that $Z_{\ph}=1/g_0^2$ at the regulator scale. If we consider
a particular quadratic superpotential ${\cal W}(\Ph)$ of the form,
\begin{equation}
{\cal W}(\Ph)=m{\rm Tr}\,\Ph^2\,,\label{perturb}
\end{equation}
which gives a bare mass 
\begin{equation}
  m_{\ph} \, = \, \frac{m}{Z_{\ph}}\, = \,g_0^2m \label{bmass}
\end{equation}
to the field $\Phi$,
then the relation (\ref{kanom1}) reduces to
\be
 \left\langle 0|\, {\rm Tr}\, \lambda^2\, |0\right\rangle \, = \,
 8\pi^2 m \, \left\langle 0|\, {\rm Tr}\, \phi^2 \, |0\right\rangle\,.
\label{kanom}
\ee

Knowledge of the gluino condensate thus boils down to knowledge of
the vacuum state in terms of $\phsq$.

\subsection{Exact Vacuum Structure}

Before reviewing the vacuum structure which emerges quantum mechanically
from the Seiberg-Witten solution, it is convenient  first to recall
the classical moduli space of ${\cal N}\!\!=\!2$ SYM. This space has 
two components,
\be
 {\cal M}_{\rm cl} \, = \, {\cal M}_{a=0} + {\cal M}_{a\neq 0},
\ee
where $a$, which we identify with the scalar vev,
$\langle 0|\ph |0 \rangle = a\, \tau_3/2$, parametrises the moduli space.
The first component, ${\cal M}_{a=0}$, is simply a point $a=0$ where the gauge
group $G=SU(2)$ remains unbroken. The second component, ${\cal M}_{a\neq 0}$,
arises for a generic vev for $\ph$ which breaks 
the gauge group from $G={\rm SU}(2)$ to $H={\rm U}(1)$, and is given
by ${\cal M}_{a\neq 0} = ({\rm \bf C}^*)^3/(G/H)_c$, 
where $(G/H)_c$ denotes the complexification of the coset, as 
is familiar in supersymmetric gauge theories. This manifold is 
simply ${\rm \bf C}^*$, the complex plane with the origin removed, 
conveniently parametrised by $a\neq 0$.
The point to emphasise here is that ${\cal M}_{\rm cl}$ is
actually two manifolds, 
the point ${\cal M}_{a=0}$ which, as we shall discuss is
associated with SCI calculations,
is distinguished by the triviality of its gauge orbit.

Turning to the quantum theory, we can focus on the low energy effective
action associated, for a generic scalar vev $a$, with the unbroken U(1).  
The light fields are described by an ${\cal N}\!\!=\!2$ U(1) vector
superfield ${\cal A}$ which contains the light multiplet associated with
the modulus $a$. 
The action is determined by the holomorphic prepotential ${\cal F}({\cal A})$,
and in terms of ${\cal N}\!\!=\!1$ superfields 
the coefficient of the U(1) gauge kinetic term $W^2$ is given by
$\,\ta(a)\equiv\, {\cal F}''(a)$, while the K\"ahler potential for the 
adjoint chiral fields determines the moduli space 
metric $ds^2=G_{a\ov{a}}|da|^2$, as $G_{a\ov{a}}={\rm Im}\,\ta/(4\pi)$. 
Classically, 
\begin{equation}
\ta_{\rm cl}=\frac{4\pi i}{g_0^2}+\frac{\th_0}{2\pi}
\label{taucl}
\end{equation}
 is constant, and the metric
is flat. However, quantum mechanically, the metric
receives a 1-loop perturbative correction, and an infinite series of 
instanton corrections~\cite{seiberg88,sw}.

While the vev $a$ is a convenient coordinate for the classical moduli space,
(and indeed can be given a gauge invariant meaning in terms of the
$W^{\pm}$--boson mass as we shall discuss later), the natural gauge
invariant modulus is given by
\be
\langle 0|\,{\rm Tr}\,\ph^2\, |0 \rangle \, =\, u\,. 
\label{udef}
\ee
The moduli  $u$ and $a$ are not independent, of course.
While classically $u=a^2/2$ and thus (up to Weyl reflections) 
each gives an equivalent parametrisation of
the classical moduli space, quantum mechanically the Seiberg-Witten 
solution~\cite{sw} determines a more
complex functional dependence and 
implies that only $u$ is a good global coordinate on the
moduli space.

We have focussed on the ${\cal N}\!\!=\!2$ moduli space geometry, since on 
perturbing
the system with a mass term for the adjoint chiral field given by 
Eq.~(\ref{perturb}), it is essentially this geometry 
which determines the scalar potential, $V$. Using the 
definition (\ref{udef}) we see that in the effective low energy
theory the perturbation (\ref{perturb}) is realised in the form
\begin{equation}
{\cal W}_{\rm eff}=m\,u
\label{supeff}
\end{equation}
and thus  the scalar  potential is given by the
reparametrisation invariant  expression,
\be
  V \, = \, \frac{|\,\ptl_z {\cal W}_{\rm eff}\,|^2}{ G_{z\ov{z}}}\, = \,
\frac{|\,m\,\ptl_z u\,|^2}{ G_{z\ov{z}}}\,. \label{V1}
\ee  
The natural choice $z=u$ for the modulus leads to
\begin{equation}
 V \, =\,|\,m\,|^2 G^{u\ov{u}}\,,
\label{V2}
\end{equation}
which exhibits the supersymmetric vacua as the singular points where the
inverse metric $G^{u\ov{u}}$ vanishes. Classically, $G^{u\ov{u}}\propto |u|$, 
i.e. the mass perturbation leads to a single vacuum state at $u_{\rm cl}=0$,
corresponding to the point ${\cal M}_{a=0}$ in the ${\cal N}\!\!=\!2$
moduli space.

Quantum mechanically, the Seiberg-Witten solution shows
that $G^{u\ov{u}}$ vanishes at two points\,\footnote{It is interesting
that at these singular points, the metric $G_{a\ov{a}}$
also vanishes, i.e. $V \propto G^{u\ov{u}}\propto G_{a\ov{a}}$.},
\begin{equation}
u\, =\,  \pm \,\La^2\,,
\label{2u}
\end{equation}
which define two ${\cal N}\!\!=\! 1$ vacua. The
dynamical scale $\La$ is defined here using Pauli-Villars 
regularisation~\cite{reg} as
\be
 \La^4 \, = \, 4 \,M_{\rm PV}^4 \exp\left(2\pi i \ta_{\rm cl}\right)\,.
\ee
Note that at $u=\infty$, which 
is the third singular point in the Seiberg-Witten 
solution, the metric $G^{u\ov{u}}$ does not vanish, instead it diverges.

To understand the appearance of singularities of the metric
at the specific points given in (\ref{2u}), recall that the 
low energy ${\cal N}\!\!=\!2$ dynamics is encoded in
an elliptic curve fibered over the $u$-plane \cite{sw}, 
\be
 y^2 \, = \, P(x,u)\,,\quad P(x,u)\, = \, (x-\La^2)(x+\La^2)(x-u)\,. 
\label{ec}
\ee
The points at which cycles of the torus can degenerate are zeros of the
discriminant,
\be
 \De = \La^2(u^2-\La^4)\,.
\ee
We can see that these zeros determine the ${\cal N}\!\!=\!1$
vacua as follows: After perturbing with
the superpotential (\ref{supeff}) let us impose the constraint (\ref{ec})
via a Lagrange multiplier $\eta$ as an addition to the superpotential,
$\Delta {\cal W}_{\rm eff}= \eta\, [y^2 - P(x,u)]$.
One then finds that for $m\neq 0$ the $F$-flatness conditions
are equivalent to the requirements: $P=0$, and $\ptl_x P=0$. The appearance 
of the second order zeros in $P(x)$ is 
equivalent to the vanishing of the discriminant (\ref{ec}). 
Thus the ${\cal N}\!\!=\!1$
vacua where $V=0$ are given by the locus of points where the torus (\ref{ec})
degenerates. We shall see another reflection of this in the form of the
scalar potential below.

Given the values in (\ref{2u}) for the modulus $u$ at the ${\cal N}\!\!=\!1$ 
vacua we obtain, by use of the Konishi anomaly in the 
form of Eq.~(\ref{kanom}), 
the gluino condensate as,
\be
 \lasq_{\rm SW} \, = \, \pm \,8\,\pi^2 m\, \La^2\,.\label{lasqsw}
\ee
Note that although we have obtained this result for small $m\ll \La$, it holds
for arbitrary $m$ due to holomorphy. The proof follows in analogy with
that in ${\cal N}=1$ SQCD \cite{hol}. We also note that
this result may be derived
without reference to the Konishi anomaly by elevating the coupling
$\ta$ to a spurion superfield, the highest component of which then
acts as a source for Tr$\la^2$ (see e.g. \cite{peskin,hklm}).

We have emphasised the role of singularities of the metric in 
fixing the ${\cal N}\!=\! 1$ vacua and thus the gluino condensate
(\ref{lasqsw}). 
At finite $m$ this approach breaks down in 
the vicinity ($\sim m/\La$) of the singularities and, as we will discuss
in Section~3, the metric is smooth in this region. However, the 
superpotential is then modified to include the contribution of
the gluino condensate, 
${\cal W}_{\rm eff}=\langle 0|{\rm Tr}\la^2|0\rangle/8\pi^2 +
mu$. This modified superpotential leads to vacua at the same points
(\ref{2u}) as can be verified, in particular, by the 
``integrating in'' procedure \cite{intin}. The result is thus self-consistent.

The sign ambiguity of the gluino condensate 
in (\ref{lasqsw}) reflects the {\bf Z}$_2$ remnant of the 
anomalous $U(1)_R$ symmetry. It is worth noting that this
{\bf Z}$_2$ symmetry suggests that, provided the potential
is smooth near $u=0$, the first derivative must also vanish at this
point, and it might be a meta-stable symmetric vacuum.
This is not the case
as may be determined by evaluating the potential (\ref{V1})
explicitly in terms of a complete elliptic 
integral $K(k)$,
\be
 V \, = \,
\frac{4\,\pi^3\,|m\,\La|^2}{|k|^2\,{\rm Re}\left(K(k'\,)\,\ov{K}(k)\right)},
  \label{potl}
\ee
where $k^2\! =\!2\La^2/(\La^2+u)$, and $(k')^2\! =\!1-k^2$. The potential 
is nonzero
at $u\! =\!0$ and actually possesses a saddle point, which is a local maximum
in ${\rm Re}\,u$ and a local minimum in Im$\,u$.

Another side remark is  that the scalar potential
(\ref{potl}) is determined by periods of the 
curve (\ref{ec}). To see this, recall that 
the modular parameter $\ta$ for
the torus (\ref{ec}) is a ratio of the two periods, $\ta\!=\!\om_1/\om_2$,
and these periods are given by $\om_1\! =\! \ptl_u {\cal F}'(a)\! 
=\!ikK(k')/\pi$
and  $\om_2 \! =\! \ptl_u a \! =\! kK(k)/\pi$. 
From (\ref{potl}), one observes that the potential is determined by
the product Im$\,(\om_1\om_2)$, and the ${\cal N}\!\!=\!1$ vacua are given 
by the 
points where either the real or imaginary parts of the periods diverge
in such a way that Im\,$\ta\rightarrow 0$,
reflecting the degeneration of one of the cycles of the elliptic
curve (\ref{ec}) as discussed above.
We will discuss the physical interpretation of the
scalar potential $V$ in more detail in Section~3.

\subsection{Strong Coupling Instanton Calculations}

We shall now compare the result (\ref{lasqsw}) obtained from the
Seiberg-Witten solution with a 1-instanton calculation using the
SCI approach. Note that this calculation is performed at
the classical vacuum, $a=0$, which is the unique SU(2) point in 
the ${\cal N}\!\!=\!2$ moduli space. Calculations with adjoint matter
fields have previously been considered in \cite{fs}.

The calculation proceeds in a manner very similar to that 
in pure ${\cal N}\!\!=\!1$ SYM, and we use the conventions 
of \cite{svrev}, to which the 
reader is referred for more details. In the pure ${\cal N}\!\!=\!1$ case, 
the instanton possesses four fermionic zero modes associated with 
$\la$, which saturate the chiral
correlator $\lasqd$. The result for this correlator is given by
\begin{equation}
\lasqd_{{\cal N}\!=\!1\, {\rm SYM}}\, = \, \frac{2^{10}}{5}\, \pi^4 
\frac{M_{\rm PV}^6}{g_0^4}\exp(2\pi i \tau_{\rm cl}).\label{eq:n1}
\end{equation}
When we add adjoint matter, there are additionally
four fermionic zero modes for $\ps$ (the superpartner of $\ph$). 
The relevant diagram is given in Fig.~1. 
\begin{figure}[h]
 \centerline{%
   \psfig{file=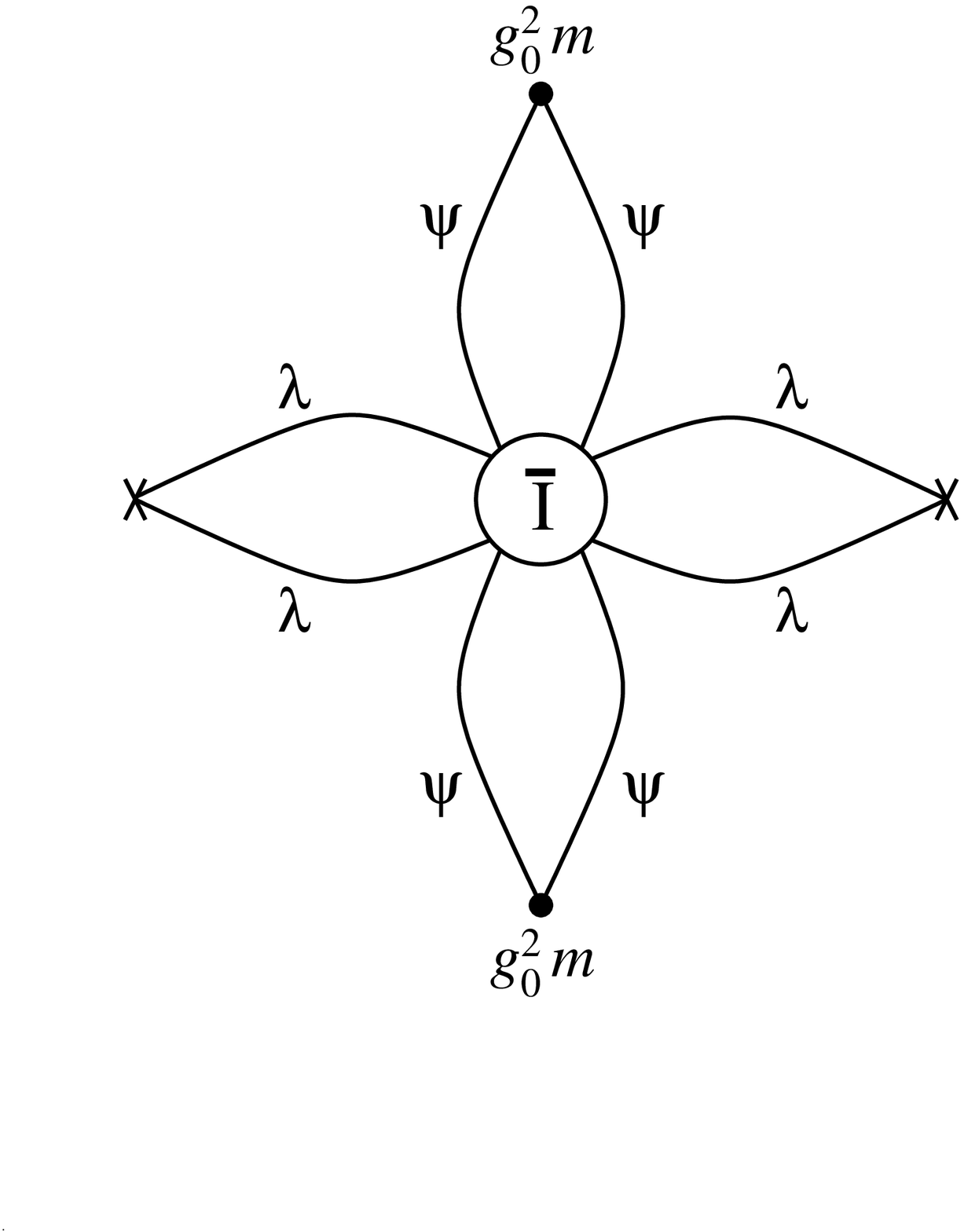,width=5cm,angle=0}%
         }
 \vskip-0.45in
\newcaption{\footnotesize
 One-(anti-)Instanton contribution to 
 $\lasqd$ in ${\cal N}\!\!=\!1$ SYM with 
 an adjoint matter field of mass $m$.}
\end{figure}
\noindent
Due to the mass perturbation, the additional matter zero-modes lead to
a factor,
\be
 \left(\frac{m_{\ph}}{M_{\rm PV}}\right)^2 \,=\, 
\left(\frac{g_0^2 m}{M_{\rm PV}}\right)^2 \,, \label{mass}
\ee
where $m_{\ph}$ is the bare mass defined in (\ref{bmass}).   
Then, it is enough to multiply Eq.~(\ref{eq:n1}) 
by the factor (\ref{mass}) to  get the result: 
\be
 \lasqd_{\rm SC} \, = \, \frac{2^{8}}{5}\, \pi^4 m^2 \La^4. \label{lasqsc}
\ee

This calculation is explicitly performed at $a=0$, which  is the classical 
vacuum in
the presence of the mass perturbation, and we find that the
result is a spacetime independent constant, as required by supersymmetry. 
It is also important to note that multi-instantons cannot contribute, 
since with $a=0$ they will not lead to the correct $m^2\La^4$ dependence.

Since the result (\ref{lasqsc}) is a constant, 
if we assume cluster-decomposition,
we find on comparison with (\ref{lasqsw}) 
a result for the gluino condensate satisfying,
\be
\left\langle 0|\, {\rm Tr}\, \lambda^2\, |0\right\rangle_{\rm SC}^2\, = \,
\frac{4}{5}\,\left\langle 0|\, {\rm Tr}\, \lambda^2\, |0\right
     \rangle_{\rm SW}^2,
\ee 
as advertised in Section~1.  The factor $4/5$ matches the discrepancy 
(\ref{disc})
observed within instanton calculations in ${\cal N}\!\!=\!1$ SQCD. 
However, in contrast with the latter theories, we have
in this case knowledge of the exact vacuum structure, and can verify that
this numerical discrepancy {\it cannot} be explained by vacuum-averaging,
which is the primary result of this paper.
In particular, there are only two chirally asymmetric 
vacuum states (\ref{2u}), each contributing equally to $\lasq^2$ according to
Eq.~(\ref{lasqsw}).

\section{SCI Calculations and the ${\cal N}\!\!=\!2$ Moduli Space}

Given the discrepancy between the SCI calculation and the Seiberg-Witten
solution uncovered in the last section, it is natural to seek an
explanation using knowledge of the exact vacuum structure. 
Indeed, its clear that the SCI calculation is performed in the
SU(2) phase which, while classically the vacuum state, is lifted quantum
mechanically. Since this lifting already occurs at the unperturbed
level of ${\cal N}\!\!=\!2$ SYM, where only the U(1) phase on Coulomb branch 
is realised, we can search for an explanation within this context.

The SU(2) phase requires the presence of a massless multiplet associated
with the $W^{\pm}$ bosons (occurring classically at $a=0$). 
Quantum mechanically these states are not present within the strong coupling
region of the moduli space, while the point $a=0$ -- where the SCI calculation
is performed\footnote{Instanton calculations have verified the 
identification of $a$ with $\langle 0| \ph | 0 \rangle$ up to 
2-instanton order in the weak coupling regime \cite{reg,2inst,yung}. 
Furthermore, a gauge invariant definition of $a$ is given by the 
(complex) mass of $W$ boson supermultiplet, whose vanishing would imply 
restoration of SU(2).
However, this is clearly true only where $W^{\pm}$ bosons exist. While
$a$ may be analytically continued into the strong coupling region, it then
has little to do with the SU(2) phase.} -- is not present. Thus
we suggest that it is the discontinuity of the spectrum across the curve
of marginal stability which is ultimately responsible for the failure
of the SCI calculation in this model. In this section we shall comment
on various aspects which lead to this conclusion.

\bigskip
\noindent
{\it The Scalar Potential}

It is worth recalling that in ${\cal N}\!\!=\!1$ SQCD with one massive
flavour, quantum corrections to the vacuum structure appear at first sight
somewhat similar to those in softly broken ${\cal N}\!\!=\!2$ SYM. 
In particular, the unique classical vacuum $a\!=\!0$ corresponds 
to the SU(2) phase, while a {\mbox 1-instanton} effect generates an 
infinite superpotential (see section~4) 
at this point, resolving the classical vacuum state to two quantum vacua.
Thus removal of the SU(2) phase is associated with the 
generation of an infinite potential at this point. 

We shall see shortly that
there are important distinctions between this scenario and
softly broken ${\cal N}\!\!=\!2$. Nonetheless, we can proceed by analogy
in considering the leading perturbative correction to the scalar 
potential $V$. This arises directly from the analogous correction to the 
moduli space metric, and has the form,
\be
 V \, = \, 4\pi^2|m|^2\frac{|a|^2}{\ln |8a^2/\La^2|}
        \left(1+O\left(\left|\frac{\La}{a}\right|^4\right)\right), 
             \label{potl2}
\ee
where instantons provide the power-like corrections.  
Importantly, this potential exhibits a singularity
at the Landau pole, beyond which, $|a|\leq |\La/2\sqrt{2}|$, it
breaks down. This singularity separates the vacuum point $a=0$
from the region $|a|\gg |\La|$ in which the potential is physically
meaningful. This presents us with a heuristic picture for the removal
of $a=0$ from the physical moduli space. However, this singular
behaviour can only be taken as a clue to what happens quantum
mechanically at  $a\sim \La$, since instanton corrections 
are numerically large in this regime, and 
manifestly singular at $a=0$, reflecting an 
induced ``tube-like'' behaviour of the metric inside the region 
$|a|\leq |\La/2\sqrt{2}|$. Nonetheless, we point out 
that the singular curve $|a|= |\La/2\sqrt{2}|$ is diffeomorphic to, and not
too far from, the quantum mechanical curve of marginal stability for 
states such as the $W^{\pm}$ bosons.

Before exploring this point in more detail, we should
emphasise that this picture is at best heuristic even including the instanton
corrections to (\ref{potl2}). In particular, 
if we consider the behaviour of $V$ near
the vacuum $u=\La^2$ in terms of the appropriate small variable
$a_D={\cal F}'(a)\ll \La$, we find a logarithmic singularity,
\be
 V \, = \,  8\pi^2\frac{|m\,\La|^2}{\ln |8\La/a_D|}
        +\cdots. \label{VD}
\ee
Of course, this singularity simply reflects a breakdown of 
the effective theory due to the presence of additional massless states
(monopoles in this case). Therefore, as noted earlier, 
although appropriate for locating the vacua, the potential we have been using 
is only valid outside a small region of the singularity of
size $O(m/\La)$. As is well known, adding the additional 
light monopole (or dyon) states, 
$\De {\cal W}_{\rm eff} = \sqrt{2}\tilde{M}a_D M$, results in smooth
vacua where these fields condense (see e.g. \cite{evans,ag,lr99} for 
explicit studies of the potential in this case). Note that the breaking
scales for {\bf Z}$_4\rightarrow\,${\bf Z}$_2$ (the discrete
remnant of U(1)$_R$), and confinement, are then related:
$\lasq = -4\sqrt{2}i\pi^2\La\langle 0| M\tilde{M}|0\rangle$.

\bigskip
\noindent
{\it The Curve of Marginal Stability}

As discussed above, the leading perturbative correction to the
potential (\ref{potl2}) is singular near the curve of marginal 
stability. Since the vacuum state at $a=0$ lies within the singular
curve, this suggests that the lifting of the SU(2) phase may 
be associated, not with a potential barrier as in ${\cal N}\!\!=\!1$ SQCD, 
but with the marginal stability of the $W^{\pm}$ boson
states on this curve. This is the point of view we shall now explore.

We begin by recalling that the
classical singular point at $u_{\rm cl}=a^2/2=0$ is degenerate in the
sense that $W^{\pm}$ bosons and monopoles both become light in the
vicinity of this point. Quantum mechanically, this singularity is
``resolved'' to the curve of marginal stability (CMS), given by
the locus $\{u |\, {\bf Im}(a_D/a)=0\}$, on which BPS states,
whose mass is given by the ${\cal N}\!\!=\!2$ central charge
\be
 M_{\, \rm BPS}\, = \, \sqrt{2}\,\bigg |\,n_e a + n_m a_D\,\bigg |\,, 
   \label{bps}
\ee
(where $n_e$, $n_m$ are integer electric and magnetic charges)
can become marginally stable. There are then only two singular points at
which charged states (monopoles and ($n_e=1,n_m=-1$) dyons) become
massless, while $W^{\pm}$ bosons remain massive.
\begin{figure}[h]
 \centerline{%
   \psfig{file=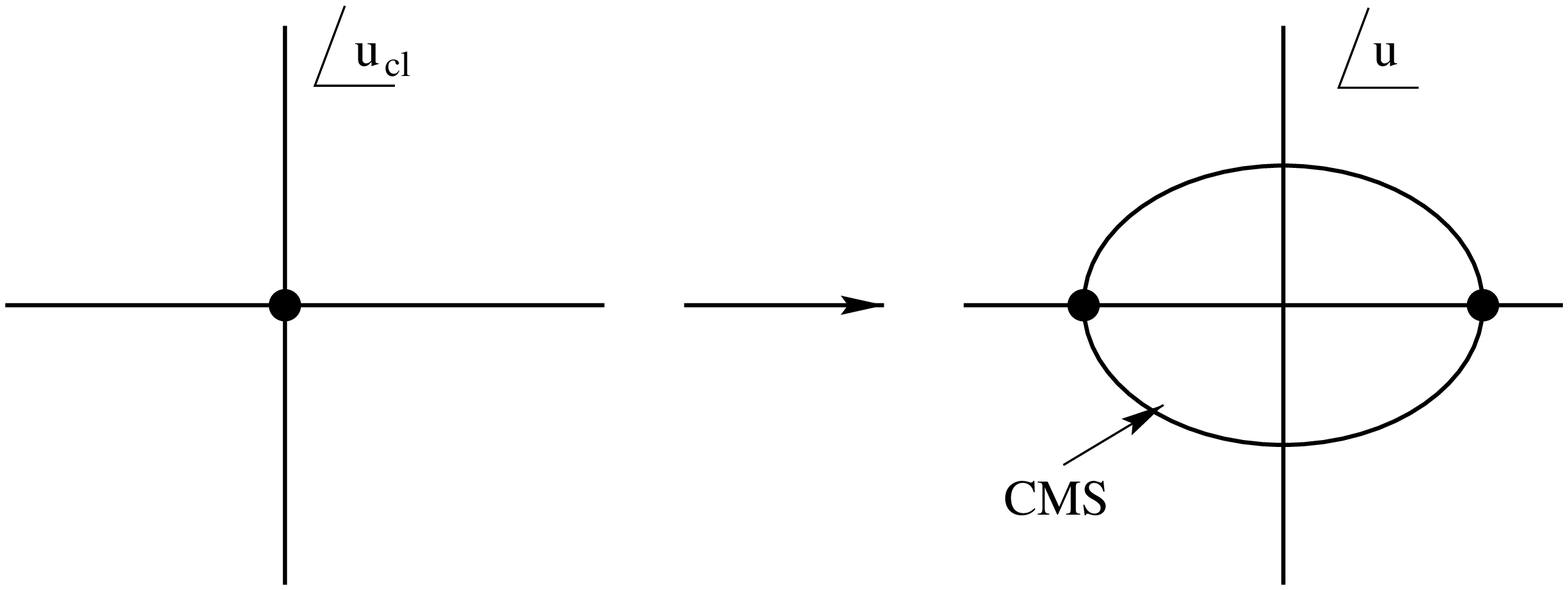,width=10cm,angle=0}%
         }
 \newcaption{\footnotesize The classical moduli space (on the left) 
$u_{\rm cl}=a^2/2$ has a 
singular point at the origin where $W^{\pm}$ bosons and monopoles are
massless. Quantum mechanically, this point is resolved to the 
CMS (on the right), where there are two distinct singularities on the
real axis.}
\end{figure}
\noindent
Indeed, while $W^{\pm}$ bosons naturally exist in the semi-classical 
spectrum, it is known that on crossing the CMS to the strong coupling
region, they no longer exist as localised one-particle 
states \cite{lr,decays}.

On the other hand, the 
relevance of the CMS curve when discussing $a=0$ is clear
when we recall that the monodromy structure of the function
$a(u)$ implies that we can formally reach the point $a_{\rm n'th-sheet}=0$
by winding round the branch cuts (in particular, the logarithmic
branch cut associated with the monopole point $u=\La^2$), 
provided we start somewhere on the first sheet where
the ratio $a_D/a$ is rational, i.e. particular points on the CMS. 
Thus Fig.~2 formally expresses the quantum mechanical
resolution of the classical point $a=0$ to a discontinuous set of points
$a_{\rm n'th sheet}=0$ lying above the CMS curve on higher Riemann sheets. 
Of course, the fact that such a point does not lie on the first sheet 
means that these points do {\it not} represent a restoration of the 
SU(2) phase. Indeed, the monodromy group is a quantum symmetry, and 
winding round to higher sheets simply moves us outside the 
fundamental domain for $a$, which amounts to a relabelling and
does not reflect a physical change.

To obtain a more physical picture for the lifting of the SU(2) phase,
we note that within the U(1) effective theory, 
the CMS itself is not singular except at the points where monopoles and 
dyons become massless. Instead it reflects a degeneracy of states. 
However, within the theory as a whole, this degeneracy
(and associated marginal stability) is reflected in a 
singularity of the $Z$--factor for massive states such as 
the $W^{\pm}$ bosons. As noted in \cite{lr}, this $Z$--factor is given by 
${\rm Im}\,({\cal F}'(a)/a)$, which  vanishes precisely on the CMS.  
The significance of this becomes apparent if we consider
trying to follow a trajectory in the moduli space from the 
semi-classical region, reducing $a$ toward zero. In doing 
so we shall necessarily reach the CMS, beyond which the $W^{\pm}$ bosons 
are not present in the spectrum \cite{decays} 
and therefore its not possible to find an SU(2) phase inside this 
region. Therefore it is natural to conclude that
the lifting of the SU(2) phase is associated physically with the
marginal quantum stability of the $W^{\pm}$ bosons
at strong coupling whose presence would be required 
to restore a full massless SU(2) multiplet. We can now interpret 
the divergence of the potential (\ref{potl2}) as a hint of 
this transition as one passes below the Landau pole.

\bigskip
\noindent
{\it The D3-Brane Probe Picture}

Given the relation between the lifting of the SU(2) phase and the
marginal stability of $W^{\pm}$ boson states that we have argued for
above, it is of interest to understand the decay of these states in
more detail. However, while this question may be addressed in certain
theories with matter, or larger gauge groups, where CMS curves are
present at weak coupling \cite{wccms}, such techniques are not 
directly applicable for CMS curves present at strong coupling, 
as is the case here.
Nonetheless, a simple geometric picture of the transition across
the CMS is available within Type IIB string theory, or more 
precisely its extension to F-theory.
Sen \cite{sen} has pointed out that the U(1) effective theory 
can be realised as the worldvolume theory on a D3-brane probe 
in the background of two mutually non-local 7-branes with the 
charges of monopoles and ($n_e=1,n_m=-1$) dyons. 
Within this framework, the BPS states present at strong coupling are
realised as strings joining the D3-brane with one of the 7-branes \cite{sen}.
However, it has been argued \cite{Wsj} that $W^{\pm}$ boson states
are realised as 4-string junctions, where the junction must lie
on the CMS for the state to be supersymmetric (see Fig.~3).   
\begin{figure}[h]
 \centerline{%
   \psfig{file=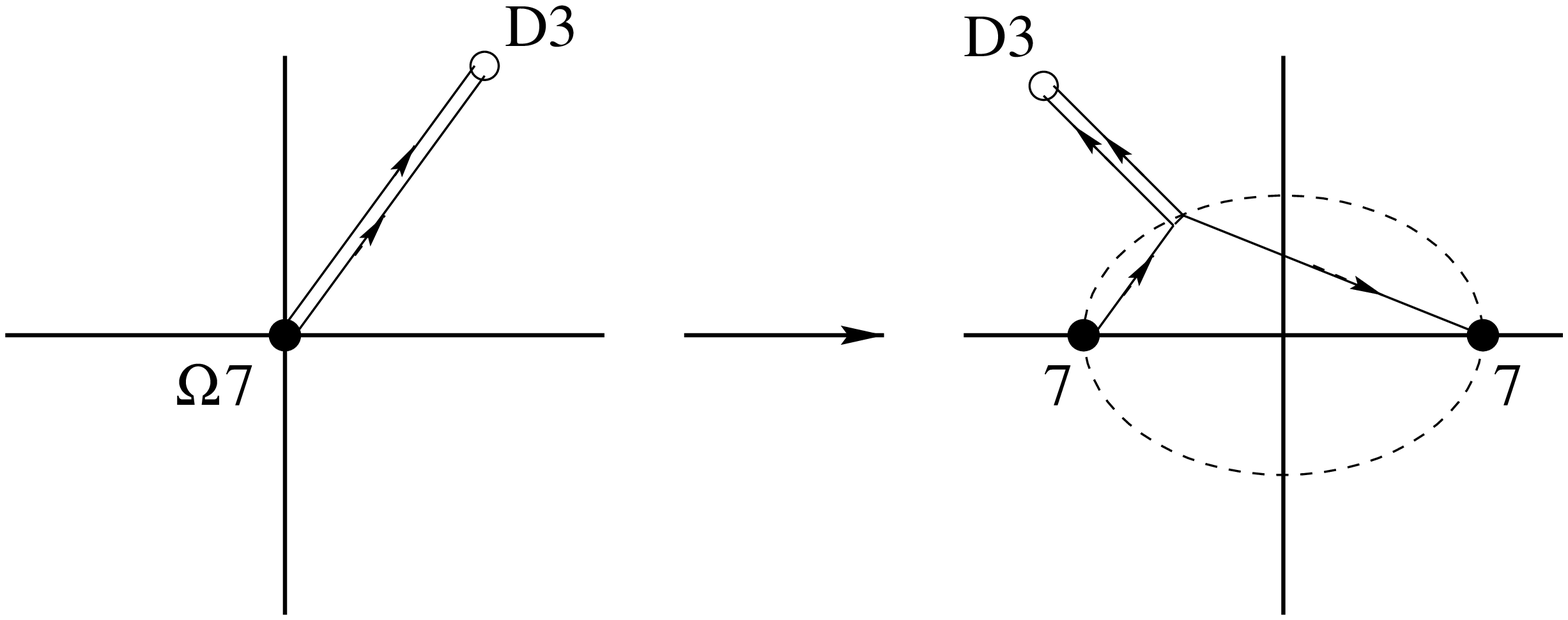,width=10cm,angle=0}%
         }
 \newcaption{\footnotesize The classical moduli space (on the left) 
with a D3-brane probe in the background of an orientifold 7-plane ($\Om 7$). 
The $W^{\pm}$ bosons are represented by an open fundamental 
string ending on the D3-brane and reflecting off the $\Om 7$-plane. 
Quantum mechanically (on the right),
the $\Om 7$-plane is resolved to two 7-branes with the respective
charges of monopoles and $(1,-1)$ dyons, and the $W^{\pm}$ bosons
are given by a particular 4-string junction configuration as shown.} 
\end{figure}
\noindent 
When the D3-brane reaches the CMS, the junction lies on the brane itself
and the the two remaining ``prongs'' can dissociate on the the D3-brane,
leaving the expected configuration of a monopole and a (1,-1) dyon, 
which is stable inside the CMS. Clearly, it would be of interest 
to understand this phenomenon within field theory.

\section{Instanton Calculations in ${\cal N}\!\!=\!1$ SQCD}

In this section, we turn to ${\cal N}\!\!=\!1$ SQCD with one flavour, and
recall a number of well-known features, 
uncovered in the 1980's \cite{wc,scmass,fs}, which 
provide additional insight into the
relation between the SCI and WCI calculations. 

SQCD exhibits some useful simplifications
with regard to instanton calculations -- namely, the 1-instanton saturation
of the chiral correlators of interest, and the mass dependence
of the vacuum state (see section~3.1). 
The model possesses chiral fields $Q_f^\alpha$, where $\alpha=1,2$ is the 
colour index  and $f=1,2$ is a sub-flavour
index. The $D$-flat directions are parametrised by a modulus superfield
$X=\sqrt{Q^{\al f}Q_{\al f}/2}$.
Classically, the addition of a mass 
perturbation, $mX^2$, leads to the presence of a single vacuum 
at $\langle 0| \varphi |0\rangle=0$ (where $\varphi$ is the lowest
component of $X$), 
i.e. the SU(2) phase, in analogy with the situation in softly 
broken  ${\cal N}\!\!=\!2$. 
However, there is also a 1-instanton induced  correction to the 
superpotential \cite{ads},
leading to an expression of the form,
\be
 {\cal W}_{\rm SQCD} \, = \, m X^2 +\frac{\La^5_1}{X^2}, \label{Wsqcd}
\ee
where $\La_1$ is a natural scale in the one flavour model,
\be
 \La_1^5 \, = \, \frac{M_{PV}^5}{g_0^4}\exp(2\pi i\ta_{\rm cl}).
      \label{La1}
\ee
We refer to bare fields
and parameters, and $Z_{\ph}(M_{PV})=1$ fixes the normalization.
Note that it s $m\,\La_1^5$ which forms a renormalisation group (RG) 
invariant, not $\La_1^5$.
This superpotential implies $\phsqb \sim \La_1^{5/2}m^{-1/2}$, 
putting the system at weak coupling in $m\ll \La_1$.

\bigskip
\noindent
{\it WCI vs SCI Calculations}

The natural chiral correlators to consider in this case are
$\lasqd$, and $m\laph$, where we recall that the 
RG invariant field
bilinears are Tr$\,\la^2$ and $m\varphi^2$. In particular, within 
the WCI approach,
the 1-instanton contribution to $\laph$ was considered
in a background with $a\gg \La$ in \cite{wc}. The instanton action
then has the form,
\be
 S_{\rm inst} \, = \, \frac{8\pi^2}{g_0^2} + 4\pi^2|a|^2\rh_{\rm inv}^2, 
                  \label{action}
\ee
where $\rh_{\rm inv}^2=\rh^2(1+4i\ov{\th}_0\ov{\beta})$ is the
supersymmetric instanton size, and $\ov{\th}_0$ and $\ov{\beta}$ 
are fermionic
collective coordinates associated respectively with supersymmetry
and superconformal transformations of the instanton (see e.g.
\cite{svrev} for details). The large vev $a$ is 
associated with the instanton induced superpotential in the
massless theory as discussed above. The result
for $\laph$ has the form
\be
 \laph \, = \, \La_1^5\, I \label{corI}\,,
\ee
and has the special feature that it is saturated by 
zero size instantons. In particular, the integral $I$ in (\ref{corI})
has the form,
\be
 I\, =\, \int \frac{{\rm d}\rh^2}{\rh^2}\,
  {\rm d}^2\ov{\th}_0{\rm d}^2\ov{\beta} 
 \, f(\rh_{\rm inv}^2) \, = \, 
16 \left[\rh^2f'(\rh^2)-f(\rh^2)\right]_0^{\infty}\,,
 \label{zero}
\ee
where the function $f$ can be written in the form \cite{wc},
\be
 f(\rh^2_{\rm inv})\,=\,\exp(-4\pi^2|a|^2\rh_{\rm inv}^2)
              \left[1-f_{\rm sc}\left(\frac{\rh_{\rm inv}^2}{(x-x')^2}\right)
            \right]\,.\label{zero1}
\ee
Here $f_{\rm sc}$ represents the contribution containing the
dependence on the $\ov{\th}_0$ and $\ov{\beta}$ zero
modes of Tr\,$\la^2$ and $\varphi^2$,
\be
 f_{\rm sc}(z)\, = \, \int_0^1 d\al\, \frac{2z^3\al^3}{[\al(1-\al)+z]^3}
    \, = \, z+O(z^2)\,,\label{zero2}
\ee
where the expansion corresponds to the limit 
$z=\rh_{\rm inv}^2/(x-x')^2 \to 0$. 

For nonzero $a$ both $f(\rh^2=\infty)$ and $f'(\rh^2=\infty)$ vanish,
and the result (\ref{zero}) for $I$  is explicitly saturated by 
zero size instantons, 
\be
I_{\rm WC}=16\, f(\rh^2=0)=16\,. 
\label{WCI}
\ee
This answer is 
determined by the instanton measure alone. The result 
displays the topological features of independence from $x-x'$,
$a$, and from the detailed structure of the instanton action
and the zero modes.
Moreover, $f_{\rm sc}$ which vanishes at $\rh^2=0$ does
not  contribute in the WCI result, which implies factorization since
the connected contribution comes only from $f_{\rm sc}$.

\begin{figure}[h]
 \centerline{%
   \psfig{file=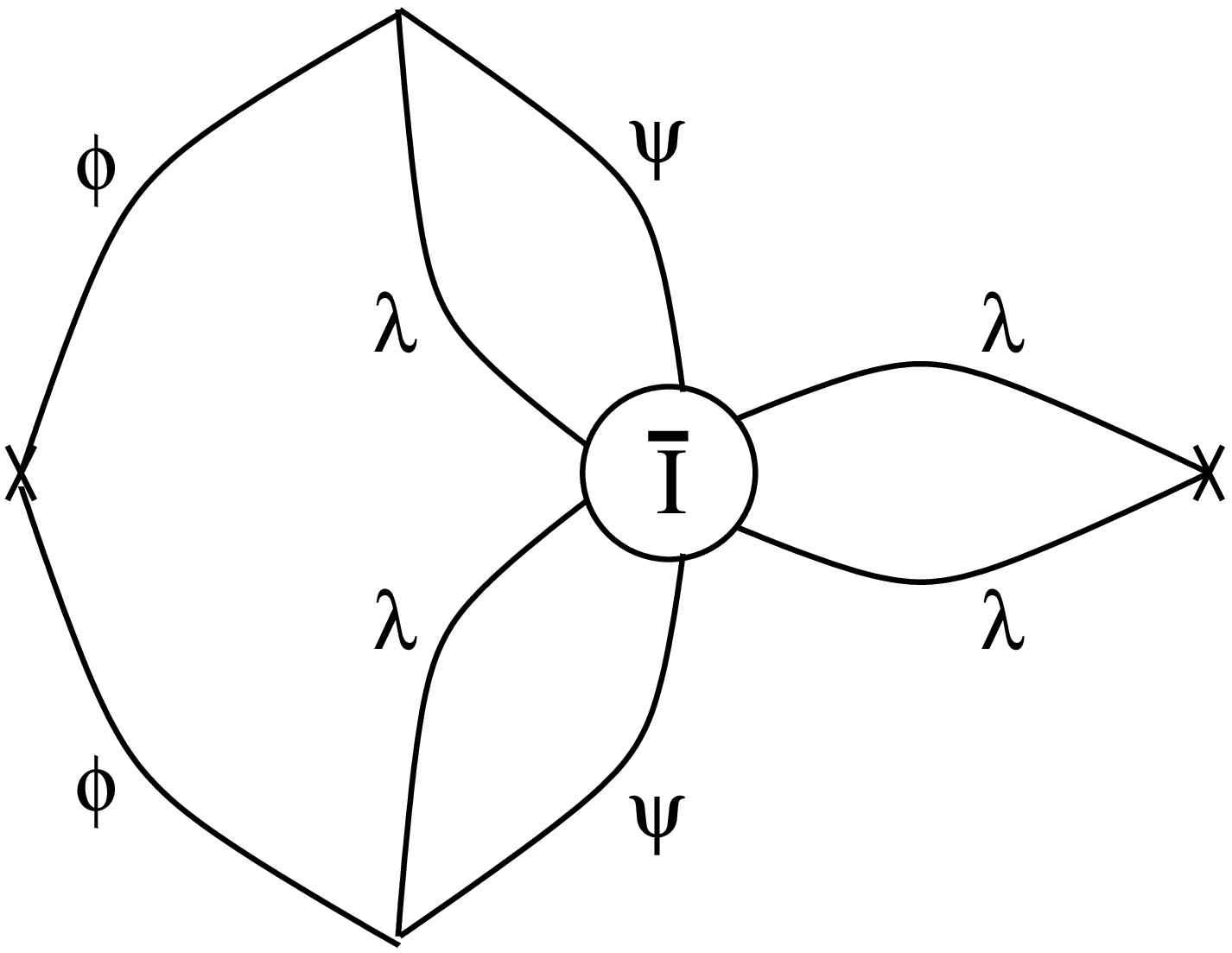,width=5.5cm,angle=0}%
         }
 \newcaption{\footnotesize One-(anti-)Instanton SCI contribution to 
$\laph$ in ${\cal N}\!\!=\!1$  SQCD.}
\end{figure}

We can use the same expressions (\ref{zero}-\ref{zero2})
to perform the analogous SCI calculation (see Fig.~4)
by setting $a=0$. Then,
\be
I_{\rm SC}=16\, f_{\rm sc}(\rh^2=\infty)=16\cdot \frac 1 2\,.
\label{SCI}
\ee
Thus we find that the WCI and SCI results 
for $\laph$ actually arise from {\it different} sources. 
While the connected part does not contribute to the WCI calculation
which is saturated by $\rho=0$, it does contribute to the SCI
calculation which is saturated at $\rho=\infty$.

It is also possible to perform the calculation of $I_{\rm SC}$ 
directly following the connected diagram given in Fig.~4 for which one finds
a result equal to (\ref{SCI}), with the distinction that it is
apparently saturated at $\rh\sim |x-x'|$. We can see that
this is in agreement with the infrared nature of the result (\ref{SCI})
by re-introducing $a\neq 0$ as an infrared regulator for 
the calculation. In this case, for any finite $a$, one observes
an exact cancellation between two integrals saturated 
respectively in the ranges $\rh\sim |x-x'|$ 
and $\rh \sim 2\pi/a$, as discussed in \cite{wc}. However, within the
SCI approach, one ignores the integral saturated at $\rh\sim 2\pi/a$
leaving a finite result as in (\ref{SCI}). Thus we see
explicitly that as we send $a\rightarrow 0$
the discontinuity arises as an infrared effect at $\rh=\infty$. 
Thus, although the details are apparently quite different, there is
an analogy with softly broken ${\cal N}\!\!=\!2$ SYM, in that
the SCI result may be interpreted in terms of unphysical 
(in this case $\rh=\infty$) configurations.

\bigskip
\noindent
{\it On the Consistency of the SCI Approach}

In addition to the arguments presented above, which serve an interpretative
role for the nature of the SCI result, this calculation also reveals an 
additional technical inconsistency associated
with the difference between the numerical factors relating the
WCI and SCI calculations for $\lasqd$ and $\laph$. 
Recall that the SCI calculation
for $\lasqd$ in ${\cal N}\!\!=\!1$ SQCD \cite{wc} is performed in analogy
with the discussion in Section~2, with the distinction that there
is only one mass insertion (see \cite{wc,svrev} for details). 
We can express the results of these calculations in the
form,
\bea
 \laph_{\rm SC} & = & \frac{1}{2}\phsqb_{\rm WC}\cdot\lasq_{\rm WC} 
                        \label{1/2}\\
 \lasqd_{\rm SC} & = & \frac{4}{5} \lasq_{\rm WC}^2. \label{4/5}
\eea
In particular, these relations imply,
\be
  \left< 0\left| {\rm Tr}\,\la^2 \left[ 16\pi^2m \varphi^2 - 
      \frac{2}{5}{\rm Tr}\,\la^2\right]\right| 0 \right> \, = \, 0.
         \label{disc2}
\ee
As we demonstrate below, this expression violates the Konishi relation
(\ref{kreln}) (due to the difference between the two coefficients in
(\ref{1/2}) and (\ref{4/5})) and is 
therefore inconsistent. This observation is not new 
(see \cite{sc,scmass,fs}), but we would like to emphasise here that 
this conclusion follows without the imposition of the assumption 
of cluster-decomposition. Thus we present the argument in some detail.

The main point is that one may
generalise the relation (\ref{kanom1}) to obtain,
\be
 \left< 0 \left|{\cal O}_1(x_1)\cdots {\cal O}_n(x_n)
\left[\frac{8\pi^2}{T_R}\, \ph\,\frac{\ptl {\cal W}(\ph)}{\ptl \ph} 
- {\rm Tr}\, \lambda^2\right]\!(y) \right| 0 \right> =0
\label{kreln2}
\ee
where ${\cal O}_i(x_i)$ are the lowest components of chiral superfields.
To prove this relation, we insert a complete set of states between
${\cal O}_n$ and the term in square brackets. The crucial point
is that this sum reduces to vacuum states as an immediate consequence
of supersymmetry -- the correlators are spacetime independent 
constants, and contributions of higher states would imply $x$-dependence.
Thus this correlator is proportional to a series of vacuum
expectation values of the form,
\be
 \left< 0_i\left| \left[\frac{8\pi^2}{T_R}\, 
    \ph\,\frac{\ptl {\cal W}(\ph)}{\ptl \ph} 
   - {\rm Tr}\, \lambda^2\right]\!(y)   \right| 0_j \right>\, = \, 0,
\ee
where $|0_i\rangle$ refers to any supersymmetric vacuum state.
This expression vanishes because, via the Konishi relation (\ref{kreln}), 
we can rewrite it as a vacuum matrix element of
a total derivative, and (\ref{kreln2}) then follows.

If we specialise (\ref{kreln2}) to the case at hand, 
for which it has the form,
\be
 \left< 0\left| {\rm Tr}\,\la^2 \left[ 16\pi^2m \varphi^2 - 
      {\rm Tr}\,\la^2\right]\right| 0 \right> \, = \, 0,
\ee
we observe a direct contradiction with (\ref{disc2}), implying as
stated above that (\ref{disc2}) is inconsistent with the Konishi
anomaly. Although this inconsistency follows trivially if we assume
cluster-decomposition, as has been noted before \cite{sc,scmass,fs},
we emphasise that we have not needed this additional
assumption, i.e. that {\it all} chiral expectation values are 
diagonal in each vacuum, 
$\langle 0_i | {\cal O} | 0_j\rangle \propto \de_{ij}$.
Of course, a direct implication is that cluster-decomposition is 
also violated. Given the discussion above which implies that the SCI result
does not contribute to the factorisable WCI calculation, this
conclusion may not be so surprising.

\bigskip
\noindent
{\it Possible Resolutions?}

A possible resolution of this discrepancy with the Konishi relation 
was suggested by Amati {\it et al.} \cite{scmass}, in that $\laph$ may have
a non-analytic mass dependence at $m=0$. An explicit calculation
in the limit $m\rightarrow \infty$ indeed verified consistency with
(\ref{kreln}), although not with the WCI results. We feel that
consistency with the Konishi relation is essentially built into this
latter calculation as one may see by viewing the massive field as a regulator.
However, since there is no indication of mass singularities of this
kind within the WCI calculation, it is not clear if this resolution
suggests a helpful physical interpretation of the numerical failure of the
SCI approach. We shall suggest a slightly different interpretation 
in Section~5.

One can ask whether the addition of a chirally symmetric vacuum state
\cite{ks} would also resolve the problem. Within this model, we have less
control than in softly broken ${\cal N}\!\!=\!2$ as the exact vacuum structure
is not strictly determined at strong coupling. An application of
cluster decomposition, with a carefully chosen weight for the 
symmetric vacuum, where $\langle 0_S | {\rm Tr}\,\la^2 |0_S\rangle= 
\langle 0_S | \varphi^2 |0_S\rangle = 0$, apparently implies that one can 
explain one of the numerical factors in (\ref{1/2}) and (\ref{4/5}) 
in this way, but not both. There is a subtlety here in 
that the Kovner-Shifman vacuum necessarily contains
massless fermions so as not to contribute to the Witten index. Thus
there is no mass-gap, and the applicability of cluster-decomposition
is not clear. However, we can use arguments based on supersymmetry as
above, to ensure that only vacuum states survive the insertion of
a complete set of states.

\bigskip
\noindent
{\it Cluster-Decomposition}

It is interesting to contrast this 1-instanton discrepancy with recent
work by Hollowood {\it et al.} \cite{hklm} on correlators saturated 
by multi-instanton configurations. Therein, calculations based on 
the $n$-instanton measure at large $N_c$, and a numerical
study of a particular SU(2) correlator saturated by 2-instanton
configurations, led the authors to conclude that ``cluster-decomposition''
was violated within the SCI approach. This failure of
cluster decomposition for particular correlators followed despite
the fact that it holds for the $n$-instanton measure itself \cite{measure}.

Note that this refers to a specific form of cluster-decomposition, 
namely the factorisation of
an $n$-instanton saturated chiral correlator into a
product of 1-instanton  saturated correlators; schematically 
represented as, 
($n$-instanton) $\rightarrow$ (\mbox{1-instanton})$^n$.
In this regard, it is worth  recalling that in simplified systems, such as 
certain 2D sigma models, where direct strong-coupling instanton 
calculations can be performed, cluster-decomposition in the
sense of multi-instanton configurations follows from the structure
of the collective coordinate integral (see e.g. \cite{sigma}). 
Thus, purely at the mathematical level, the underlying reason for
its apparent failure in SCI calculations
in ${\cal N}\!\!=\!1$ SYM remains something of a mystery. We note only
that as we emphasised earlier the classical
moduli space is disconnected. Thus it may be more appropriate to consider
2D sigma models on disconnected target spaces.

In contrast, while the discrepancy we have discussed above
also implies a breakdown of cluster-decomposition in a simple sense, 
it is quite distinct from these multi-instanton considerations.
Specifically, we only
consider 1-instanton effects, and thus cluster decomposition
represented in the form $\laph\rightarrow \phsqb\lasq$ cannot be interpreted
in terms of a factorisation of instanton contributions. Here the
notion is more basic in that, as described above, the WCI result
factorises trivially as it is effectively a calculation of
$\lasq$ in a background field.

\section{Concluding Remarks}

In this note we have discussed various inconsistencies
of the strong coupling approach to the calculation of 
instanton induced chiral correlators. In particular, we have argued
that within a specific model -- softly broken ${\cal N}\!\!=\!2$ SYM -- 
the vacuum
averaging hypothesis cannot be used to explain the numerical 
value of $\lasq_{\rm SC}$. Furthermore, within ${\cal N}\!\!=\!1$ SQCD with one
flavour, although we have no clear picture of the vacuum structure at strong
coupling in this case, we recalled that an apparent
technical inconsistency at 1-instanton order with the Konishi relation
is still perfectly consistent with the factorisability of the weak 
coupling calculation as the connected part associated with 
SCI actually vanishes.

Since are not strictly at liberty to integrate
out the adjoint chiral field in softly broken ${\cal N}\!\!=\!2$, we have
little to say about the possible existence (or non-existence) of the chirally
symmetric Kovner-Shifman vacuum in pure ${\cal N}\!\!=\!1$ SYM. In particular,
for consistency, we have always assumed when working with
an adjoint chiral field that $m\ll \La$. For $m\geq \La$ there may be
additional states to consider, 
and we are not guaranteed that vacuum rearrangements will not take place.
However, we note that the arguments of Kovner and Shifman in favor of
the new vacuum rely heavily on the vacuum averaging hypothesis
\cite{screv} for SCI calculations.

The inconsistencies we have discussed 
suggest that the SCI approach 
is deficient with regard to extracting physical correlators in the 
specific ${\cal N}\!\!=\!1$ theories studied in this paper. In particular,
within softly broken 
${\cal N}\!\!=\!2$, it is apparent that the SCI calculation, saturated
at 1-instanton order, is referred to the SU(2) phase which, while a
classical vacuum, is lifted quantum mechanically. In contrast, the
result arising from the Seiberg-Witten solution, is instead
interpreted as an effect due to an infinite series of instanton
corrections within the U(1) phase. Thus the discrepancy does not
seem so surprising, and this
raises the question of whether the SCI result has a well-defined
interpretation. One possibility follows from the fact that 
the chiral correlators studied here apparently
define particular topological invariants~\footnote{Since we work in flat
space, this is manifest only in independence from the spacetime
coordinates. Extension
to specific nontrivial geometries does however indicate dependence
only on the intrinsic topology \cite{svunp}.}. 
Indeed they bear a close relation to expectation values 
calculable on generic backgrounds
within Witten's twisted ${\cal N}\!\!=\!2$ formulation of 
Donaldson theory. However,
this picture applies quite generally to the correlators themselves, and is
not specific to a particular instanton approximation. Nonetheless, given 
that the SCI approach determines unambiguous numerical values for these
invariants, one may wonder whether they are related to a particular
topological variant of ${\cal N}\!\!=\!1$ SYM.

\subsection*{Acknowledgments}
We would like to thank A.~Gorsky, I.~Kogan, A.~Kovner,
M.~Shifman, and A.~Yung for helpful discussions.
This work was supported in part by 
the Department of Energy under Grant No. DE-FG02-94ER40823.


\end{document}